\pdfoutput=1
\documentclass[11pt,onecolumn,a4paper]{article}

\usepackage{gensymb, graphicx, amsmath, amssymb,fullpage}
\usepackage{sectsty}
\allsectionsfont{\normalfont\sffamily\bfseries}

\title{\sffamily\bfseries Optimal incentives for collective intelligence}

\author{Richard P. Mann$^{1*}$ and Dirk Helbing$^2$ \\ 
{\small 1. Dept. of Statistics, School of Mathematics, University of Leeds, Leeds LS2 9JT, UK.} \\
{\small 2. Computational Social Science, ETH Zurich, Zurich 8092, Switzerland.}\\
{\small * Email: r.p.mann@leeds.ac.uk}}

\date{}

\begin{document}
\maketitle

\begin{abstract}
Collective intelligence is the ability of a group to perform more effectively than any individual alone. Diversity among group members is a key condition for the emergence of collective intelligence, but maintaining diversity is challenging in the face of social pressure to imitate one's peers. We investigate the role incentives play in maintaining useful diversity through an evolutionary game-theoretic model of collective prediction. We show that market-based incentive systems produce herding effects, reduce information available to the group and suppress collective intelligence. In response, we propose a new incentive scheme that rewards accurate minority predictions, and show that this produces optimal diversity and collective predictive accuracy. We conclude that real-world systems should reward those who have demonstrated accuracy when majority opinion has been in error.
\end{abstract}

The financial crisis and its aftermath have reopened long-standing debates about the collective wisdom of our societal organisations \cite{galton1907vox,mackay2012extraordinary,hertwig2012tapping}. Financial and prediction markets seem unable to foresee major economic and political upheavals such as the credit crunch, Brexit and the American presidential election. This lack of collective foresight might be the result of insufficient diversity among decision-making individuals \cite{shefrin2009psychological}. Diversity is a crucial condition for collective intelligence \cite{surowiecki2005wisdom,page2008difference,zafeiris2013gpi,page2014wdc, aplin2014ilp} that can be more important than the intelligence of individuals within a group \cite{Woolley2010efa}. As collective intelligence ultimately results from individual actions, it depends on how individuals are incentivised \cite{pickard2011time,hong}. While most previous research has focused on explaining how the phenomena of collective intelligence emerge \cite{couzin2009collective}, less is known about how to optimise collective wisdom in a quantitative sense. 

Harnessing collective wisdom is important. Global systems of communication, governance, trade and transport grow rapidly in complexity every year. As a result it becomes impossible for any single individual or agency to gather and process enough data to understand the entire system \cite{helbing2015saving}. Attention is therefore shifting towards decentralised systems as a means to bring together the local knowledge and private expertise of many individuals \cite{lammer2008self,pickard2011time}. In machine-learning, researchers have found that diverse ensembles of models maximise prediction accuracy \cite{bell2007lessons}. In politics, the forecasts of prediction markets \cite{wolfers2006pmi, arrow2008promise} are now commonly reported alongside opinion polls during elections. Scientists are also turning to crowd-sourcing collective wisdom as a validation tool \cite{oprea2009crowdsourcing,morgan2014use, herbert2015turing}. However, as highlighted by the inability of financial and prediction markets to foresee the results of recent elections in the UK and USA, collective wisdom is not a guaranteed property of a distributed system \cite{mackay2012extraordinary}, in part due to herding effects \cite{lorenz2011hsi,moussaid2013opinion}. In science as well, the incentive structure undervalues diversity: low-risk projects with assured outcomes are more likely to be funded than highly novel or interdisciplinary work \cite{young2008current,stephan2012research}. Rewards for conformity with institutional cultures can severely limit useful diversity as well \cite{duarte2015political}. Previous work \cite{prelec2004bayesian} has investigated mechanisms to elicit truthful, existing minority views to counter herding effects in expressed opinion. This raises the question: how can minority viewpoints be fostered in the first place, so as to enhance diversity and its potential benefits for collective intelligence?

Here we analyse an evolutionary game-theoretic model of collective intelligence amongst unrelated agents motivated by individual reward. We show that previously proposed incentive structures \cite{hong} are sub-optimal from the standpoint of collective intelligence, and in particular produce too little diversity between individuals. We propose a new incentive system we term `minority rewards', wherein agents are rewarded for expressing accurate minority opinions, and show that this produces stable, near-optimal collective intelligence at equilibrium. Our results demonstrate that existing real-world reward structures are unlikely to produce optimal collectively intelligent behaviour, and we present a superior alternative that should motivate new reward systems.

\section*{Results}
To investigate the effect of incentives on collective intelligence, we use an abstract model of collective information gathering and aggregation \cite{hong}. Complex outcomes are modelled as a result of $n$ independent, causal factors. A large population of individual agents gather information in a decentralised fashion, each being able to pay attention to just one of these factors at any given time. Collective prediction is achieved by aggregation of individual predictions via simple voting. Agents are motivated to seek information and to provide predictions by incentive schemes that offer rewards for making accurate predictions. It is assumed that the accuracy of an individual's prediction can be judged after the event. We exclude cases where the ground truth is either never discoverable or where no such ground truth exists (for instance in questions regarding taste or voter preferences), but instead consider questions such as predictions of future events (which are known once they occur) or scientific questions (which may be resolved at some later time). To illustrate with concrete examples, one might consider whether national GDP will rise above trend in the coming year, whether a given company will increase its profits, or whether global temperatures will increase by more than 1\degree C in the next decade. The proportion of agents attending to different sources of information evolve depending on the rewards they receive, where less successful agents tend to imitate their more successful peers. 

Consider a binary outcome, $Y$, which is the result of many factors, $x_1, x_2, \ldots, x_n$. We model this outcome as the sign of a weighted sum of the contributing factors:
\begin{equation}
Y = \textrm{sign}\left(\sum_{i=1}^n \beta_i x_i \right).
\end{equation}
Each contributing factor takes binary values, such that $Y, x_i \in \{-1,1\}$. We assume that the values of these factors are uncorrelated (see Supplementary Information for instances where this assumption may be relaxed). Without loss of generality we assume that $\beta_i > 0$ for all factors.

An individual attending to factor $i$ observes the value of $x_i$. Having observed the value of $x_i$, this individual then votes in line with that observation. Thus, if the proportion of individuals attending to factor $i$ is $\rho_i$, the collective prediction, $\hat{Y}$ is given by:
\begin{equation}
\hat{Y} = \textrm{sign}\left(\sum_{i=1}^n \rho_i x_i \right).
\end{equation}
Collective accuracy, $C$, is the probability that the collective vote agrees with the ground truth, given the distribution, $\{\rho\}$, of agents attending to each factor:
\begin{equation}
C = P\left(\hat{Y} = Y \mid \{\rho\}\right).
\end{equation}
The reward given to an agent for an accurate vote depends on the proportion of other correct votes in any given collective decision. We denote as $z_i$ the proportion of agents that will vote identically to those attending to factor $i$, that is the proportion of agents attending to factors whose value matches $x_i$: $z_i = \sum_{j=1}^n \rho_j \delta_{x_i, x_j}$, where $\delta$ is the Kronecker delta. The reward is determined by a function, $f(z_i)$, such that an agent receives a reward proportional to $f(z_i)$ if and only if their prediction is accurate. We will investigate three potential reward systems for deciding how each agent is rewarded for their accurate votes, the first two of which are taken from previous work by Hong \emph{et al.} \cite{hong}. The first of these is `binary rewards': agents receive a fixed reward if they make an accurate prediction, corresponding to the reward function $f(z_i) = 1$. The second is `market rewards': a fixed total reward is shared equally amongst all agents who vote accurately, corresponding to the reward function $f(z_i) = 1/z_i$. This adds an incentive to be accurate when others are not, and closely mimics the reward system of actual prediction markets. Finally we introduce `minority rewards': agents are rewarded for an accurate prediction when fewer than half of the other agents also vote accurately, corresponding to the reward function $f(z_i) = 1-H(z_i-1/2)$, where $H(\cdot)$ is the Heavyside step function. This explicitly rewards agents who hold accurate \emph{minority} opinions, and incentivises agents to be accurate on questions where the aggregate prediction is wrong.

The expected reward a player receives by attending to factor $i$ is determined by the expected value of $f(z_i)$, conditioned on voting accurately (Methods, equation \ref{eqn:expected_reward}).Players adapt their behaviour in response to the rewards they and others receive. In alignment with previous evolutionary game theory work, we model changes in individual attention to factors as being the result of imitation; agents who are observed to be gaining greater rewards are imitated by those gaining fewer. This leads to the classic replicator equation \cite{helbing1996stochastic}, describing the evolution of the proportion of agents, $\rho_i$, that pay attention to factor $i$ (Methods, equation \ref{eqn:replicator})

We studied the behaviour of the model under the three incentive schemes described above. We initialised the model by assigning uniform proportions of agents to each factor, with values of $\beta$ randomly drawn from a uniform distribution (the absolute scale of $\beta$ does not affect the model). We followed the evolutionary dynamics described by the replicator equation until the population converged to equilibrium. This was repeated over a range of problem dimensionalities from $n=3$ to $n=10000$. Expected rewards were calculated either by exhaustive search over all possible values of $x_1, \ldots, x_n$ (for $n < 10$) or by using appropriate normal-distribution limits for large numbers of factors (see Methods). 
\begin{figure}[h!]
\centering
\includegraphics[width=0.9\linewidth]{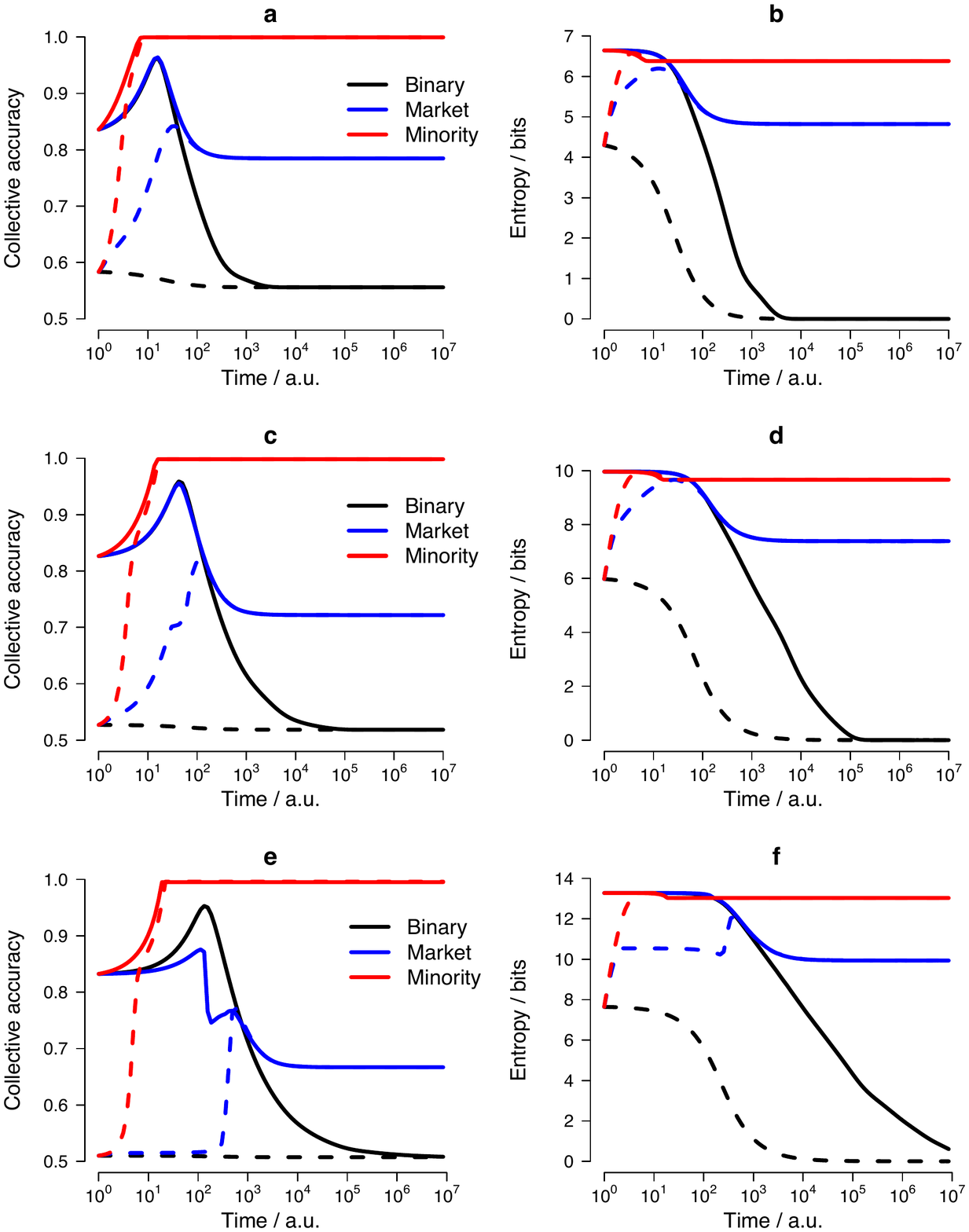}
\caption{Evolution of collective accuracy (a,c,e) and diversity (b, d, f) for binary rewards (black line), market rewards (blue line) and minority rewards (red line) in simulations with $n=100$ (a,b), $n=1000$ (c,d) and $n=10,000$ (e,f) independent factors. Solid lines indicate results from a uniform initial allocation of agents over factors, while dashed lines indicate an initial allocation of 50\% of agents to the single most important factor, with the remainder allocated uniformly over the remaining factors. Note that the number of time steps is plotted on a logarithmic scale}
\label{fig:approach_equilibrium}
\end{figure}
Figure \ref{fig:approach_equilibrium} shows how collective accuracy and diversity evolve towards equilibrium for the three rewards systems of binary, market and minority rewards in simulations with $n=100$, $n=1000$ and $n=10,000$ independent factors. Note the logarithmic scale on the x-axis, to better illustrate the early evolution. For each reward system two initial allocations of agents' attention are used: (i) a uniform allocation to each factor ; and (ii) an allocation where half of all agents attend to the single most important factor, with others allocated uniformly across the other factors. This demonstrates that the equilibrium distribution of attention is the same whether agents initially attend to arbitrary factors or initially favour the most obvious ones. The exact time scale of convergence to equilibriums depends on the magnitude of rewards; in our simulations we normalise rewards such that the mean reward \emph{per agent} is one at each time step.

\begin{figure}[!h]
\centering
\includegraphics[width=0.6\linewidth]{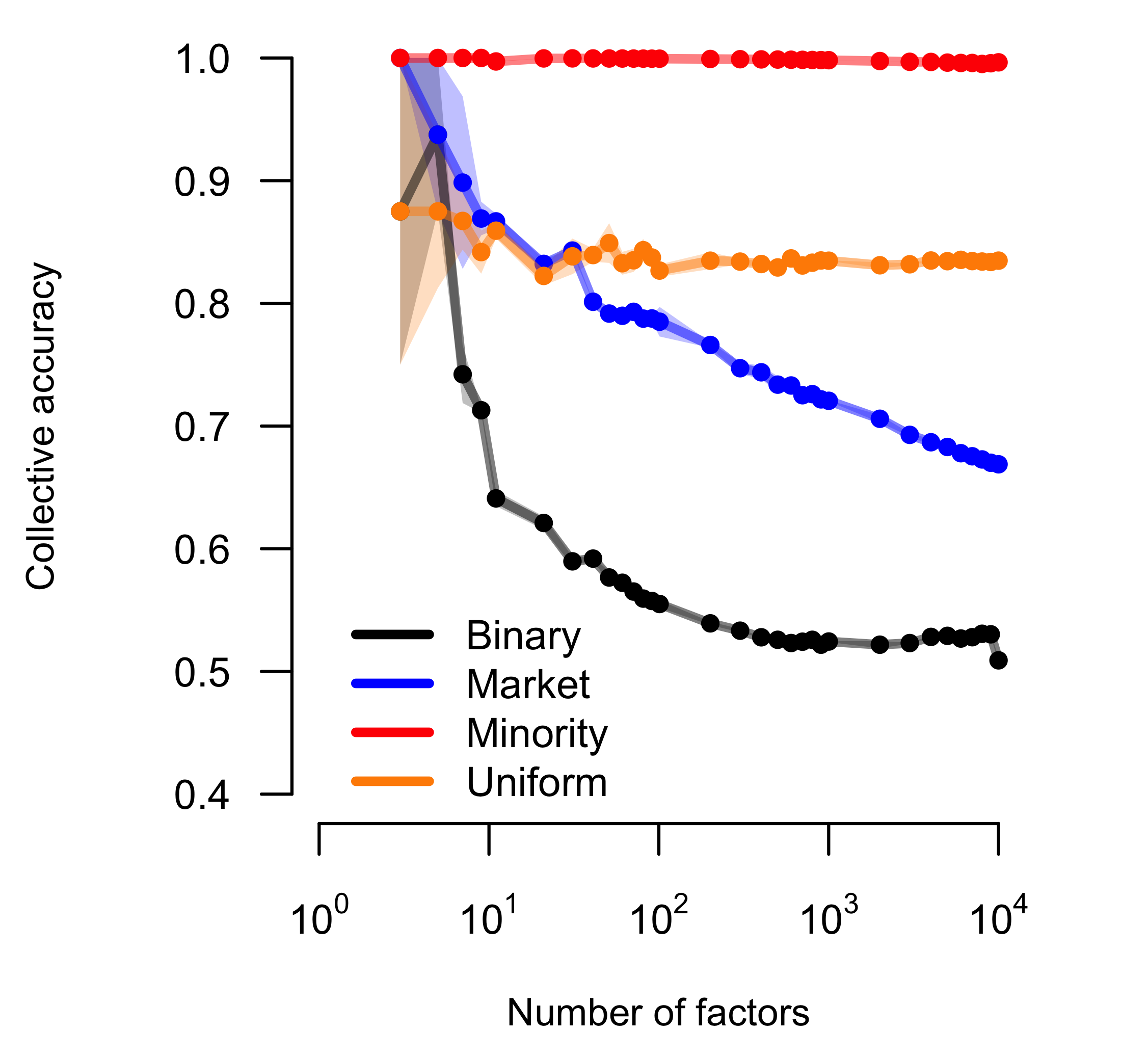}
\caption{\textbf{Collective accuracy at equilibrium as a function of the number of independent factors across different reward systems. Solid lines and shaded regions show the mean and standard deviation from 10 independent simulations with different randomly generated values for the factor coefficients. Points on each curve show the precise values of $n$ for which simulations were carried out, equally spaced within each multiple of 10.}}
\label{fig:1}
\end{figure}
Figure \ref{fig:1} shows how the resulting collective accuracy varies across problem dimensionalities from $n=3$ to $n=10000$ for the three different reward systems and for a uniform allocation of agents to factors. Consistent with \cite{hong}, we find that market rewards increase diversity and collective accuracy relative to binary rewards. However, collective accuracy under market rewards declines rapidly with increasing $n$, falling to $\sim 65\%$ for $n=10000$. For comparison we also show the accuracy achieved under a uniform allocation of agents, which reaches a stable value of approximately 80\% for large $n$. Market rewards therefore produce lower accuracy than a uniform allocation for all but the lowest values of $n$. In contrast, minority rewards lead to a far higher accuracy than any of the investigated alternative reward systems regardless of system complexity, and achieve close to 100\% accuracy up to $n = 10000$. Our mathematical analysis shows that minority rewards will continue to produce near-perfect accuracy for any problem size, as long as the population of agents remains large relative to the number of factors (see Supplementary Information).

\begin{figure}[!h]
\centering
\includegraphics[width=0.9\linewidth]{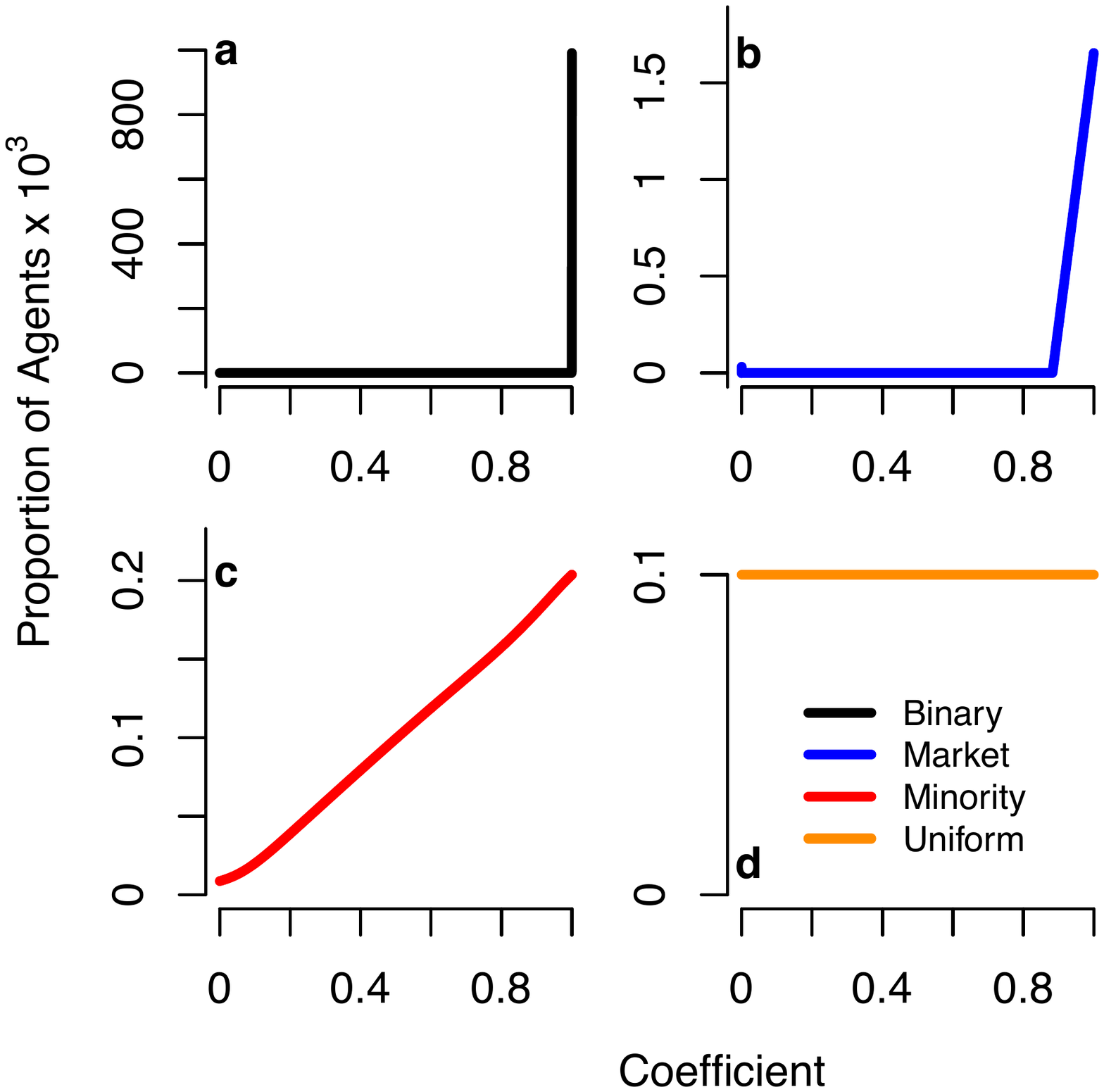}
\caption{\textbf{Equilibrium proportions of agents paying attention to each factor, as a function of the coefficient associated with that factor. Results are shown for simulations with $n=10000$ factors, and for the three reward systems of binary rewards (a), market rewards (b) and minority rewards (c), as well as the uniform allocation (d). Binary rewards drive almost all agents to the single most important factor (the greatest coefficient). Market rewards create a distribution proportional to coefficient size across the most important 10\% of factors, while minority rewards distribute agents almost perfectly in proportion to the magnitude of the coefficient.}}
\label{fig:2}
\end{figure}
The different levels of collective accuracy across reward systems are a reflection of the differing equilibrium distributions of the proportion of agents attending to each factor. The minority rewards scheme outperforms both market rewards and uniform, unweighted approaches, as it automatically redirects attention if the aggregate prediction would otherwise be wrong. Under minority rewards the system converges towards a state where the number of agents paying attention to any factor is proportional to its importance. This optimal distribution is both a stationary and stable state of the minority rewards system (see Supplementary Information). In Figure \ref{fig:2} we plot the equilibrium distribution for each reward system for a high-dimensional problem ($n=10000$). Using binary rewards, almost all agents attend to the single most important factor. Under market rewards agents distribute themselves in proportion to the predictive value of the factors, but only amongst the top 10\% of factors; 90\% of factors receive essentially no attention at all. By comparison, under minority rewards the proportion of agents paying attention to a factor is also proportional to its importance, but agents cover the full range of factors down to the least important, providing more information to the group and improving predictions. 

\section*{Discussion}
We constructed a novel reward system, minority rewards, that incentivises individual agents in their choice of which informational factors to attend to when operating as part of a group. This new system rewards agents both for making accurate predictions and for being in the minority of their peers or conspecifics. As such it encourages a balance between seeking useful information that has substantive predictive value for the ground truth, and seeking information that is currently under utilised by the group. We evaluated the accuracy of collective prediction within our model resulting from our new reward scheme against  both previously proposed market-like reward mechanisms and the maximally diverse, uniform allocation of agents attention. Our results and analysis show that minority rewards induces optimal collective intelligence, while market rewards have lower performance that deteriorates with group size.

The poor performance of market rewards relative to a uniform unweighted allocation for $n > 10$ shows that a market reward system incentivises herding behaviour and suppresses useful diversity, as illustrated by the equilibrium distribution in Figure \ref{fig:2}b. This suggests that stock markets and prediction markets may systematically underweight a large pool of informational factors that are of limited predictive power individually, but which can contribute powerfully to aggregate predictions if agents can be persuaded to pay attention to them. This sheds doubt on the accuracy of existing markets as a tool for aggregating dispersed knowledge to predict future profits or events, and motivates further work on designing collectively more accurate market mechanisms. The relatively high performance of uniform allocations of attention supports work showing that linear models with equally weighted predictors can match or even improve on more closely fitted prediction models \cite{dawes1979robust,graefe2015improving}. Inclusion of all relevant predictors is often more important than determining their appropriate weights in making predictions; too much diversity is less harmful than too little, especially for complex problems. 

Incentives are a fundamental part of any effort to harness the potential of collective intelligence. In this paper we have presented evidence that rewarding accurate minority opinions induces near-optimal collective accuracy. Therefore, to maximise the collective wisdom of a group, we suggest that individuals should not be rewarded simply for having made successful predictions or findings, nor should a total reward be equally distributed amongst those who have been successful or accurate. Instead, rewards should be primarily oriented towards those who have made successful predictions in the face of majority opposition from their peers, i.e. towards those who tell us something we don't already know. Future work should investigate how such a reward system could be implemented in practice, to improve career progression, funding and reputation systems \cite{conte2012manifesto}, prediction markets, and democratic procedures \cite{helbing2016we}. In conclusion, therefore, how best to foster collective intelligence is an important problem we need to solve collectively.

\section*{Methods and Materials}

\subsection*{Terminology}
Throughout this paper we use the following conventions for describing probability distributions:
\begin{itemize}
\item $\mathbb{E}(x)$ denotes the expectation of $x$
\item $\mathcal{N}(x; \mu, \sigma^2)$ denotes the normal probability density function with mean $\mu$ and variance $\sigma^2$, evaluated at $x$
\item $\mathcal{N}(x; \mu, \Sigma)$ for vector-valued $x$ and $mu$, and matrix $\Sigma$ denotes the multi-variate normal probability density function with mean $\mu$ and covariance matrix  $\Sigma$, evaluated at $x$
\item $\Phi(x)$ denotes the standard normal cumulative probability distribution function with mean 0 and standard deviation 1.
\end{itemize}

\subsection*{Ground truth and voting}
We consider a binary outcome, $Y$ that is the result of many independent factors, $x_1, x_2, \ldots, x_n$. We model this outcome as being determined by the sign of $\psi$: a weighted sum of the contributing factors. To facilitate further analysis we introduce the latent variable $\psi$, which is the simple sum of the contributions from each factor.
\begin{equation}
Y = \textrm{sign}(\psi), \ \psi = \sum_{i=1}^n \beta_i x_i.
\end{equation}
In computational implementation of this model we sample values of $\{\beta\}$ independently from a uniform distribution (the scale of which is arbitrary and does not influence the analysis). We assume without loss of generality that factors are ordered such that $\beta_i \geq \beta_{i+1}$, and further we normalise the values of the coefficients such that $\sum_{i=1}^n \beta_i = 1$, without affecting the value of $Y$. Our analytical results below do not depend on the exact distributon of $\{\beta\}$. Any sampling distribution for $\{\beta\}$ that has a finite moment of order $m, \ m > 2$ will obey the Ljapunov and Lindeberg conditions \cite{feller1970prob}, guaranteeing convergence in distribution of $\psi$ to a normal distribution, from which our results are obtained.

Each individual attends to one factor at a given time; an individual attending to factor $i$ therefore observes the value of $x_i$. Having observed the value of $x_i$ this individual then votes in line with that observation. The collective prediction, $\hat{Y}$ is given by the sign of the collective vote $V$, which is a sum over the contributing factors, weighted by the proportion of individuals attending to each factor:
\begin{equation}
\hat{Y} = \textrm{sign}(V), \ V = \sum_{i=1}^n \rho_i x_i.
\end{equation}

\subsection*{Evolutionary dynamics}
We model changes in individual attention to factors as being motivated by imitation individuals; agents who are observed to be gaining greater rewards are imitated by those gaining fewer, leading to the classic replicator equation \cite{helbing1996stochastic} describing the evolution of $p_i$, the proportion of agents attending to factor $i$:
\begin{equation}
\dot{\rho_i} = \rho_i \left( \mathbb{E}(R_i) - \sum_{j=1}^n \rho_j \mathbb{E}(R_j) \right) \label{eqn:replicator},
\end{equation}
where $\sum_{i=1}^n \rho_i = 1$ by definition. When solving these $n$ equations (one for each factor) numerically, we normalise the total rewards given to all agents such that $\sum_{i=1}^n \rho_i \mathbb{E}(R_i) = 1$. This is equivalent to adaptive variation of the time step and does not change the relative rewards between options, nor the final steady state, but ensures smoother convergence to that state. This also mimics a real constraint on any practical reward system where the total reward available may be fixed. In our simulation of the collective dynamics of the system we used the Runge-Kutta order 2(3) algorithm, as implemented in R by Soetaert \emph {et al.} \cite{soetaert2010solving}.
\\\\
\subsection*{The three reward schemes}
In the main text we present three possible systems for rewarding agents for making accurate predictions. Each reward scheme corresponds to a choice of reward modulation function, $f(z)$, which determines the magnitude of the reward when an agent makes an accurate prediction, as a function of the proportion, $z$, of other agents that also do so. These are:
\begin{enumerate}
\item Binary rewards: $f(z) = 1$
\item Market rewards: $f(z) = 1/z$
\item Minority rewards: $f(z) = 1-H(z-1/2)$, where $H$ is the Heavyside step-function. 
\end{enumerate}
The expected reward an agent receives for attending to factor $i$ is therefore the expected value of $f(z_i)$, conditional on their vote being accurate:  
\begin{equation} 
\mathbb{E}(R_i) = \int_{\epsilon}^1 f(z_i) P(Y = x_i \mid z_i)p(z_i) dz.
\end{equation}
where $z_i$ is the proportion of agents voting identically to those attending to factor $i$: $z_i = \sum_{j=1}^n \rho_j \delta_{x_i, x_j}$, where $\delta$ is the Kronecker delta. The lower limit of the integral above is $\epsilon > 0$ to account for the limiting case of a single individual attending to the factor. As the population size $N$ tends to infinity, $\epsilon$ tends to zero. For our implementation we take $\epsilon = 10^{-6}$.

\subsection*{Normal approximation for expected rewards} 
In our model individuals are rewarded for making correct predictions. Rewards vary according to how many other individuals are correct. Focusing on a single individual who attends to factor $i$, we can calculate the expected reward received by the individual as follows. Firstly, we assume without loss of generality by symmetry that the focal individual observes $x_i = 1$. We denote as $z_i$ the proportion of individuals who vote the same way as the focal individual, and allow for a general reward modulation function $f(z)$. With these conditions, the expected reward, $\mathbb{E}(R_i)$ is:
\begin{equation} 
\mathbb{E}(R_i) = \int_0^1 f(z_i) P(\psi > 0 \mid x_i=1, z_i) p(z_i)dz \label{eqn:expected_reward}
\end{equation}
For a relatively small number of independent factors, $n$ (we use $n<10$) this integral can be evaluated by exhaustive enumeration of all possible values of $x_1, \ldots, x_n$, and thus all possible values of $\psi$ and $z_i$. Otherwise, assuming $n$ to be large, we use normal approximations to the relevant probabilities in this integral. 

Given the independence of the individual values of $x_i$, the mean and variance of $\psi$ can be determined by the linearity of expectations and by the sum rule for variances of independent variables:
\begin{equation}
\begin{split}
\mathbb{E}(\psi \mid x_i = 1) &= \beta_i \sum_{j \neq i}^n \beta_j \mathbb{E}(x_j) = \beta_i \\
\textrm{\textsc{var}}(\psi \mid x_i = 1) &=\sum_{j \neq i}^n \beta_j^2 \mathbb{E}(x_j^2) = \sum_{j \neg i}^n \beta_j^2 \\
\Rightarrow p(\psi \mid x_i = 1) &\simeq \mathcal{N}\left(\psi; \beta_i, \sum_{j \neq i} \beta_j^2\right)
\end{split}
\end{equation}
In the case of binary rewards, where $f(z) = 1$, the value of $z_i$ does not impact on the reward for attending to any factor. In this case the expected reward is calculated directly from the distribution of $\psi$:
\begin{equation} 
\begin{split}
\mathbb{E}_{\textrm{binary}}(R_i) &= P(\psi > 0 \mid x_i=1)\\
&= \int_0^\infty \mathcal{N}\left(\psi; \beta_i, \sum_{j \neq i} \beta_j^2\right) d\psi \\
&= \Phi \left(\frac{\beta_i} {\sum_{j \neq i} \beta_j^2}\right)
\end{split}
\end{equation}

For other reward schemes where the value of $z$ affects the reward, we also require an approximation for $p(z_i)$. Again we calculate the mean and variance of $z_i$:
\begin{equation}
\begin{split}
\mathbb{E}(z_i \mid x_i=1) &= \rho_i  + \sum_{j \neq i} \rho_j (\mathbb{E}(x_j)+1)/2 = (1+\rho_i)/2 \\
\textrm{\textsc{var}}(z_i \mid x_i=1) &=\sum_{j \neq i}^n (\rho_j/2)^2 \mathbb{E}(x_j^2) = \frac{1}{4}\sum_{j \neq i} \rho_j^2 \\
\Rightarrow p(z_i \mid x_i=1) &\simeq \mathcal{N}\left(z; \frac{1+\rho_i}{2}, \frac{1}{4}\sum_{j \neq i} \rho_j^2\right)
\end{split}
\end{equation}
The convergence of $z_i$ in distribution to a normal distribution depends on the values of $\{\rho\}$ meeting the Lindeberg condition \cite{feller1970prob}. In practice this means that all elements of $\{\rho\}$ should tend to zero as the number of dimensions, $n$ tends to infinity, i.e. the distribution should not be dominated by a small subset of elements. As illustrated in Figure \ref{fig:approach_equilibrium}, when the system is initialised in a state conforming to these requirements it will remain so for market and minority reward systems, but not for the binary reward system. Since the binary reward system does not depend on the value of $z_i$ the failure of this approximation in this case does not have any repurcussions for our results.

$\psi$ and $z_i$ are correlated due to the shared dependence on the values of $x_1, \ldots, x_n$, with a covariance of:
\begin{equation}
\begin{split}
\textrm{\textsc{cov}}(z_i, \psi \mid x_i =1) &= \frac{1}{2}\sum_{j \neq i} \sum_{k \neq i} \beta_j\rho_k \mathbb{E}(x_jx_k)\\
& = \frac{1}{2}\sum_{j \neq i}\beta_j\rho_j
\end{split}
\end{equation}
In the normal distribution limit,the joint distribution may be approximated as
\begin{equation}
p(\psi, z_i \mid x_i = 1) = \mathcal{N} \left( \begin{bmatrix} \psi \\ z_i \end{bmatrix}; \begin{bmatrix} \mu_{\psi} \\ \mu_z \end{bmatrix}, \begin{bmatrix} K_{\psi, \psi}  & K_{\psi, z} \\ K_{\psi, z} & K_{z,z} \end{bmatrix} \right)
\end{equation}
with,
\begin{align*}
\mu_{\psi} &= \mathbb{E}(\psi \mid x_i=1)\\
\mu_z &= \mathbb{E}(z_i \mid x_i=1)\\
K_{\psi, \psi} &= \textrm{\textsc{var}}(\psi \mid x_i=1)\\
K_{z,z} &= \textrm{\textsc{var}}(z_i \mid x_i=1)\\
K{\psi,z} &= \textrm{\textsc{cov}}(\psi, z_i \mid x_i=1) 
\end{align*}
Using standard relations for conditional normal distributions we therefore have:
\begin{equation}
\begin{split}
p(\psi \mid x_i=1, z_i) &= \mathcal{N} \left(\psi; \mu_{\psi} + (z_i - \mu_z)\frac{K_{\psi, x}}{K_{z,z}}, K_{\psi, \psi}- \frac{K_{\psi, x}^2}{K_{z,z}} \right)\\
\Rightarrow P(\psi > 0 \mid x_i=1, z) &= \Phi\left(\frac{\mu_{\psi} + (z_i - \mu_z)\frac{K_{\psi, x}}{K_{z,z}}} {K_{\psi, \psi}- \frac{K_{\psi, x}^2}{K_{z,z}}} \right)
\end{split}
\end{equation}
Combining the above expressions gives the complete equation for the expected reward of attending to factor $i$, conditioned on the values of $\beta$, the current distribution of attention, $\rho$, and the reward function $f(z)$
\begin{equation}
\mathbb{E}(R_i) = \int_\epsilon^1 f(z_i) \mathcal{N}\left(z_i; \mu_z, K_{z,z} \right)\Phi\left(\frac{\mu_{\psi} + (z_i - \mu_z)\frac{K_{\psi, x}}{K_{z,z}}} {K_{\psi, \psi}- \frac{K_{\psi, x}^2}{K_{z,z}}} \right)dz_i
\end{equation}
This integral may be evaluated numerically to give the expected reward for any general reward modulation function $f(z)$.

\subsection*{Calculating collective accuracy}
The collective accuracy, $C$, is the probability that the collective vote will correctly predict the ground truth, conditioned on the current distribution of attention to different factors. For small numbers of factors (we use $n < 10$) this can be determined exactly by exhaustive search over all $2^n$ possible combinations of the values of $x_1, \ldots x_n$. For larger values of $n$ we use the following normal approximation (similarly defined as above) for the joint distribution of the latent ground truth function $\psi$ and the collective vote $V$. 
\begin{equation}
p(\psi, V) \simeq \mathcal{N}\left(\begin{bmatrix} \psi \\ V\end{bmatrix}; \begin{bmatrix} 0 \\ 0 \end{bmatrix}, \begin{bmatrix} S_{\psi, \psi}  & S_{\psi, V} \\ S_{\psi, V} & S_{V,V} \end{bmatrix} \right)
\end{equation}
where
\begin{equation}
S_{\psi, \psi} = \sum_{i=1}^n \beta_i^2, \ 
S_{V, V} = \frac{1}{4}\sum_{i=1}^n \rho_i^2, \ 
S_{\psi, V} = \frac{1}{2}\sum_{i=1}^n \beta_i \rho_i,
\end{equation}
implying the following conditional probability distribution for $V$ given $\psi$:
\begin{equation}
p(V \mid \psi) \simeq \mathcal{N}\left(V; \psi \frac{S_{\psi, V}}{S_{\psi, \psi}}, S_{V, V} - \frac{S_{\psi, V}^2}{S_{\psi, \psi}}  \right).
\end{equation}
Considering without loss of generality the case where $Y = 1$,
\begin{equation}
\begin{split}
C &= P(\hat{Y} = 1 \mid Y = 1) \\
&= P(V > 0 \mid \psi > 0) \\
&= 2\int_{0}^\infty \int_0^\infty \mathcal{N}\left(V; \psi \frac{S_{\psi, V}}{S_{\psi, \psi}}, S_{V, V} - \frac{S_{\psi, V}^2}{S_{\psi, \psi}}  \right) dV \mathcal{N}\left(\psi; 0, S_{\psi, \psi}\right) d\psi \\
 &= 2\int_{0}^\infty \Phi\left(\frac{\psi \frac{S_{\psi, V}}{S_{\psi, \psi}}}{ S_{V, V} - \frac{S_{\psi, V}^2}{S_{\psi, \psi}} } \right) dV \mathcal{N}\left(\psi; 0, S_{\psi, \psi}\right) d\psi,
\end{split}
\end{equation}
which can be evaluated numerically. By maximising the above expression with respect to the distribution of $\rho$, one can show the intuitive result that collective accuracy is maximised when $\rho \propto \beta$, at which point collective accuracy is 100\%. The normal approximation limit becomes invalid when the distribution of $\{\rho\}$ is concentrated on very few elements; in these cases (which we identify as 99\% of the distribution mass being concretrated on fewer than 10 elements) we use exhaustive search over the values of $\{x\}$ corresponding to the remaining factors with a non-negligible values of $\rho$.

\subsection*{Acknowledgements}
We would like to thank Cedric Beaume, Viktoria Spaiser and Jochen Vo{\ss} for helpful discussions and comments on the manuscript. This work was supported by ERC Advanced Investigator Grant `Momentum', reference number 324247.

\bibliographystyle{naturemag}

\begin{thebibliography}{10}
\expandafter\ifx\csname url\endcsname\relax
  \def\url#1{\texttt{#1}}\fi
\expandafter\ifx\csname urlprefix\endcsname\relax\def\urlprefix{URL }\fi
\providecommand{\bibinfo}[2]{#2}
\providecommand{\eprint}[2][]{\url{#2}}

\bibitem{galton1907vox}
\bibinfo{author}{Galton, F.}
\newblock \bibinfo{title}{Vox populi (the wisdom of crowds)}.
\newblock \emph{\bibinfo{journal}{Nature}} \textbf{\bibinfo{volume}{75}},
  \bibinfo{pages}{450--451} (\bibinfo{year}{1907}).

\bibitem{mackay2012extraordinary}
\bibinfo{author}{Mackay, C.}
\newblock \emph{\bibinfo{title}{Extraordinary popular delusions and the madness
  of crowds}} (\bibinfo{publisher}{Start Publishing LLC},
  \bibinfo{year}{2012}).

\bibitem{hertwig2012tapping}
\bibinfo{author}{Hertwig, R.}
\newblock \bibinfo{title}{Tapping into the wisdom of the crowd -- with
  confidence}.
\newblock \emph{\bibinfo{journal}{Science}} \textbf{\bibinfo{volume}{336}},
  \bibinfo{pages}{303--304} (\bibinfo{year}{2012}).

\bibitem{shefrin2009psychological}
\bibinfo{author}{Shefrin, H.}
\newblock \bibinfo{title}{How psychological pitfalls generated the global
  financial crisis}.
\newblock \emph{\bibinfo{journal}{Insights into the Global Financial Crisis}}
  \bibinfo{pages}{224} (\bibinfo{year}{2009}).

\bibitem{surowiecki2005wisdom}
\bibinfo{author}{Surowiecki, J.}
\newblock \emph{\bibinfo{title}{{The Wisdom of Crowds}}}
  (\bibinfo{publisher}{Random House LLC}, \bibinfo{year}{2005}).

\bibitem{page2008difference}
\bibinfo{author}{Page, S.~E.}
\newblock \emph{\bibinfo{title}{{The Difference: How the Power of Diversity
  Creates Better Groups, Firms, Schools, and Societies}}}
  (\bibinfo{publisher}{Princeton University Press}, \bibinfo{year}{2008}).

\bibitem{zafeiris2013gpi}
\bibinfo{author}{Zafeiris, A.} \& \bibinfo{author}{Vicsek, T.}
\newblock \bibinfo{title}{Group performance is maximized by hierarchical
  competence distribution}.
\newblock \emph{\bibinfo{journal}{Nature Communications}}
  \textbf{\bibinfo{volume}{4}} (\bibinfo{year}{2013}).

\bibitem{page2014wdc}
\bibinfo{author}{Page, S.~E.}
\newblock \bibinfo{title}{Where diversity comes from and why it matters?}
\newblock \emph{\bibinfo{journal}{European Journal of Social Psychology}}
  \textbf{\bibinfo{volume}{44}}, \bibinfo{pages}{267--279}
  (\bibinfo{year}{2014}).

\bibitem{aplin2014ilp}
\bibinfo{author}{Aplin, L.~M.}, \bibinfo{author}{Farine, D.~R.},
  \bibinfo{author}{Mann, R.~P.} \& \bibinfo{author}{Sheldon, B.~C.}
\newblock \bibinfo{title}{Individual-level personality influences social
  foraging and collective behaviour in wild birds}.
\newblock \emph{\bibinfo{journal}{Proc. Roy. Soc. B}}
  \textbf{\bibinfo{volume}{281}}, \bibinfo{pages}{20141016}
  (\bibinfo{year}{2014}).

\bibitem{Woolley2010efa}
\bibinfo{author}{Woolley, A.~W.}, \bibinfo{author}{Chabris, C.~F.},
  \bibinfo{author}{Pentland, A.}, \bibinfo{author}{Hashmi, N.} \&
  \bibinfo{author}{Malone, T.~W.}
\newblock \bibinfo{title}{Evidence for a collective intelligence factor in the
  performance of human groups}.
\newblock \emph{\bibinfo{journal}{Science}} \textbf{\bibinfo{volume}{330}},
  \bibinfo{pages}{686--688} (\bibinfo{year}{2010}).

\bibitem{pickard2011time}
\bibinfo{author}{Pickard, G.} \emph{et~al.}
\newblock \bibinfo{title}{Time-critical social mobilization}.
\newblock \emph{\bibinfo{journal}{Science}} \textbf{\bibinfo{volume}{334}},
  \bibinfo{pages}{509--512} (\bibinfo{year}{2011}).

\bibitem{hong}
\bibinfo{author}{Hong, L.}, \bibinfo{author}{Page, S.~E.} \&
  \bibinfo{author}{Riolo, M.}
\newblock \bibinfo{title}{Incentives, information, and emergent collective
  accuracy}.
\newblock \emph{\bibinfo{journal}{Managerial and Decision Economics}}
  \textbf{\bibinfo{volume}{33}}, \bibinfo{pages}{323--334}
  (\bibinfo{year}{2012}).

\bibitem{couzin2009collective}
\bibinfo{author}{Couzin, I.~D.}
\newblock \bibinfo{title}{Collective cognition in animal groups}.
\newblock \emph{\bibinfo{journal}{{Trends in Cognitive Sciences}}}
  \textbf{\bibinfo{volume}{13}}, \bibinfo{pages}{36--43}
  (\bibinfo{year}{2009}).

\bibitem{helbing2015saving}
\bibinfo{author}{Helbing, D.} \emph{et~al.}
\newblock \bibinfo{title}{Saving human lives: what complexity science and
  information systems can contribute}.
\newblock \emph{\bibinfo{journal}{{Journal of Statistical Physics}}}
  \textbf{\bibinfo{volume}{158}}, \bibinfo{pages}{735--781}
  (\bibinfo{year}{2015}).

\bibitem{lammer2008self}
\bibinfo{author}{L{\"a}mmer, S.} \& \bibinfo{author}{Helbing, D.}
\newblock \bibinfo{title}{Self-control of traffic lights and vehicle flows in
  urban road networks}.
\newblock \emph{\bibinfo{journal}{Journal of Statistical Mechanics: Theory and
  Experiment}} \textbf{\bibinfo{volume}{2008}}, \bibinfo{pages}{P04019}
  (\bibinfo{year}{2008}).

\bibitem{bell2007lessons}
\bibinfo{author}{Bell, R.~M.} \& \bibinfo{author}{Koren, Y.}
\newblock \bibinfo{title}{{Lessons from the Netflix prize challenge}}.
\newblock \emph{\bibinfo{journal}{ACM SIGKDD Explorations Newsletter}}
  \textbf{\bibinfo{volume}{9}}, \bibinfo{pages}{75--79} (\bibinfo{year}{2007}).

\bibitem{wolfers2006pmi}
\bibinfo{author}{Wolfers, J.} \& \bibinfo{author}{Zitzewitz, E.}
\newblock \bibinfo{title}{Prediction markets in theory and practice}.
\newblock \bibinfo{type}{Working Paper} \bibinfo{number}{12083},
  \bibinfo{institution}{National Bureau of Economic Research}
  (\bibinfo{year}{2006}).
\newblock \urlprefix\url{http://www.nber.org/papers/w12083}.

\bibitem{arrow2008promise}
\bibinfo{author}{Arrow, K.~J.} \emph{et~al.}
\newblock \bibinfo{title}{The promise of prediction markets}.
\newblock \emph{\bibinfo{journal}{Science}} \textbf{\bibinfo{volume}{320}},
  \bibinfo{pages}{877--878} (\bibinfo{year}{2008}).

\bibitem{oprea2009crowdsourcing}
\bibinfo{author}{Oprea, T.~I.} \emph{et~al.}
\newblock \bibinfo{title}{{A crowdsourcing evaluation of the NIH chemical
  probes}}.
\newblock \emph{\bibinfo{journal}{Nature Chemical Biology}}
  \textbf{\bibinfo{volume}{5}}, \bibinfo{pages}{441--447}
  (\bibinfo{year}{2009}).

\bibitem{morgan2014use}
\bibinfo{author}{Morgan, M.~G.}
\newblock \bibinfo{title}{Use (and abuse) of expert elicitation in support of
  decision making for public policy}.
\newblock \emph{\bibinfo{journal}{Proceedings of the National Academy of
  Sciences}} \textbf{\bibinfo{volume}{111}}, \bibinfo{pages}{7176--7184}
  (\bibinfo{year}{2014}).

\bibitem{herbert2015turing}
\bibinfo{author}{Herbert-Read, J.~E.}, \bibinfo{author}{Romenskyy, M.} \&
  \bibinfo{author}{Sumpter, D.~J.}
\newblock \bibinfo{title}{A turing test for collective motion}.
\newblock \emph{\bibinfo{journal}{Biology Letters}}
  \textbf{\bibinfo{volume}{11}}, \bibinfo{pages}{20150674}
  (\bibinfo{year}{2015}).

\bibitem{lorenz2011hsi}
\bibinfo{author}{Lorenz, J.}, \bibinfo{author}{Rauhut, H.},
  \bibinfo{author}{Schweitzer, F.} \& \bibinfo{author}{Helbing, D.}
\newblock \bibinfo{title}{How social influence can undermine the wisdom of
  crowd effect}.
\newblock \emph{\bibinfo{journal}{Proceedings of the National Academy of
  Sciences}} \textbf{\bibinfo{volume}{108}}, \bibinfo{pages}{9020--9025}
  (\bibinfo{year}{2011}).

\bibitem{moussaid2013opinion}
\bibinfo{author}{Moussa{\"\i}d, M.}
\newblock \bibinfo{title}{Opinion formation and the collective dynamics of risk
  perception}.
\newblock \emph{\bibinfo{journal}{{PLoS ONE}}} \textbf{\bibinfo{volume}{8}},
  \bibinfo{pages}{e84592} (\bibinfo{year}{2013}).

\bibitem{young2008current}
\bibinfo{author}{Young, N.~S.}, \bibinfo{author}{Ioannidis, J.~P.} \&
  \bibinfo{author}{Al-Ubaydli, O.}
\newblock \bibinfo{title}{Why current publication practices may distort
  science}.
\newblock \emph{\bibinfo{journal}{PLoS Med}} \textbf{\bibinfo{volume}{5}},
  \bibinfo{pages}{e201} (\bibinfo{year}{2008}).

\bibitem{stephan2012research}
\bibinfo{author}{Stephan, P.~E.}
\newblock \bibinfo{title}{Research efficiency: Perverse incentives}.
\newblock \emph{\bibinfo{journal}{Nature}} \textbf{\bibinfo{volume}{484}},
  \bibinfo{pages}{29--31} (\bibinfo{year}{2012}).

\bibitem{duarte2015political}
\bibinfo{author}{Duarte, J.~L.} \emph{et~al.}
\newblock \bibinfo{title}{Political diversity will improve social psychological
  science}.
\newblock \emph{\bibinfo{journal}{Behavioral and Brain Sciences}}
  \textbf{\bibinfo{volume}{38}}, \bibinfo{pages}{e130} (\bibinfo{year}{2015}).

\bibitem{prelec2004bayesian}
\bibinfo{author}{Prelec, D.}
\newblock \bibinfo{title}{{A Bayesian truth serum for subjective data}}.
\newblock \emph{\bibinfo{journal}{Science}} \textbf{\bibinfo{volume}{306}},
  \bibinfo{pages}{462--466} (\bibinfo{year}{2004}).

\bibitem{helbing1996stochastic}
\bibinfo{author}{Helbing, D.}
\newblock \bibinfo{title}{A stochastic behavioral model and a `microscopic'
  foundation of evolutionary game theory}.
\newblock \emph{\bibinfo{journal}{{Theory and Decision}}}
  \textbf{\bibinfo{volume}{40}}, \bibinfo{pages}{149--179}
  (\bibinfo{year}{1996}).

\bibitem{dawes1979robust}
\bibinfo{author}{Dawes, R.~M.}
\newblock \bibinfo{title}{The robust beauty of improper linear models in
  decision making.}
\newblock \emph{\bibinfo{journal}{{American Psychologist}}}
  \textbf{\bibinfo{volume}{34}}, \bibinfo{pages}{571} (\bibinfo{year}{1979}).

\bibitem{graefe2015improving}
\bibinfo{author}{Graefe, A.}
\newblock \bibinfo{title}{Improving forecasts using equally weighted
  predictors}.
\newblock \emph{\bibinfo{journal}{Journal of Business Research}}
  \textbf{\bibinfo{volume}{68}}, \bibinfo{pages}{1792--1799}
  (\bibinfo{year}{2015}).

\bibitem{conte2012manifesto}
\bibinfo{author}{Conte, R.} \emph{et~al.}
\newblock \bibinfo{title}{Manifesto of computational social science}.
\newblock \emph{\bibinfo{journal}{The European Physical Journal Special
  Topics}} \textbf{\bibinfo{volume}{214}}, \bibinfo{pages}{325--346}
  (\bibinfo{year}{2012}).

\bibitem{helbing2016we}
\bibinfo{author}{Helbing, D.}
\newblock \bibinfo{title}{Why we need democracy 2.0 and capitalism 2.0 to
  survive}.
\newblock \emph{\bibinfo{journal}{Jusletter IT}}  (\bibinfo{year}{2016}).

\bibitem{feller1970prob}
\bibinfo{author}{Feller, W.}
\newblock \emph{\bibinfo{title}{An Introduction to Probability Theory and Its
  Applications: Volume 1}}, vol.~\bibinfo{volume}{1} (\bibinfo{publisher}{John
  Wiley \& Sons}, \bibinfo{year}{1970}), \bibinfo{edition}{3} edn.

\bibitem{soetaert2010solving}
\bibinfo{author}{Soetaert, K.}, \bibinfo{author}{Petzoldt, T.} \&
  \bibinfo{author}{Setzer, R.~W.}
\newblock \bibinfo{title}{{Solving differential equations in R: package
  deSolve}}.
\newblock \emph{\bibinfo{journal}{Journal of Statistical Software}}
  \textbf{\bibinfo{volume}{33}}, \bibinfo{pages}{1--25} (\bibinfo{year}{2010}).

\end{thebibliography}

\clearpage

\section*{Supplementary Information}
\section*{Mathematical analysis of stability of stationary states}

\subsection*{Stationary solution for the binary rewards scheme}
Under binary rewards agents receive a fixed reward for voting correctly. Under these conditions the stationary and stable distribution is one with all agents attending to a single factor, the one with the greatest value of $\beta$. The expected reward for agents in this system is simply proportional to the probability of voting correctly. If we set the fixed reward to be 1, and assuming without loss of generality that a focal agent observes factor $x_i = 1$ this can be written as:
\begin{equation}
\begin{split}
\mathbb{E}(R_i) &= P(\psi > 0 \mid x_i = 1) \\
&= P(\beta_i + \sum_{j \neq i} \beta_jx_j > 0).
\end{split}
\end{equation}
Since $\sum_{j \neq i} \beta_jx_j$ has an expectation of zero and a variance that decreases with increasing $\beta_i$ the above expression is clearly a monotonically increasing function of $\beta_i$. Agents that observe factors other than that with the highest value of $\beta$ can always improve their expected reward by moving to the factor with the highest value of $\beta$. As such the distribution with all agents attending to this factor is both stationary and stable.

\subsection*{Stationary solution for the minority rewards scheme}
The ideal distribution froma standpoint of collective accuracy is: $\rho_i = \beta_i$, recalling that we normalise the values of $\{\beta\}$ such that $\sum_{k=1}^n \beta_k=1$. This distribution is intrinsically stationary under the minority rewards system. If $\rho_i = \beta_i$ then the collective vote is always accurate, implying that $P(\psi > 0 \mid x_i, z_i) = 0 \ \forall z_i < 0.5$, which implies that no reward is possible for $z_i < 0.5$. At the same time the reward structure implies that no reward is possible for $z_i > 0.5$, so the reward for attending to any factor is zero. Therefore there is no evolutionary pressure for agents to change the factor that they attend to.

\subsection*{Stability of the ideal distribution for the minority rewards scheme}
We analyse the stability of the stationary ideal distribution by considering the evolution of small perturbations away from this stationary state. The expected reward as a function of $\{\rho\}$ is non-differentiable at the stationary point, so we cannot perform standard linear stability analysis for arbitrary pertubations. Instead we consider two special types of perturbation: (i) a small transfer of agent desnsity from one factor, $j$ to another, $i$, leaving all other factors unchanged, and (ii) an extensive perturbation over all factors, such that the distribution of perturbations over factors obeys the Lindeberg condition \cite{feller1970prob} and thus permits limiting approximations involving the normal distribution.
\\\\
\noindent \textbf{Pertubation on two factors}. First we consider a pertubation of the form:
\begin{equation}
\begin{split}
\rho_i &= \beta_i + \Delta, \ \Delta > 0\\
\rho_j &= \beta_j - \Delta \\
\rho_k &= \beta_k, \ \forall k \neq i,j
\end{split}
\end{equation}
In the case of such a pertubation we need to consider the rewards received by agents attending to each factor under four scenarios, based on the values of $x_i$ and $x_j$: (i) $x_i=x_j =1$; (ii) $x_i=x_j=-1$; (iii) $x_i = 1, x_j=-1$; (iv) $x_i = -1, x_j = 1$. Each of these scenarios occurs with probability $1/4$. We will address each in turn, considering the expected reward for attending to factor $i$, factor $j$ and a general factor $l, \, l \neq i, j$.
\\\\
\noindent \textbf{Scenario 1}, $x_i=1, x_j=1$:
The value of $\psi$:
\begin{equation}
\begin{split}
(\psi \mid x_i = 1, x_j = 1) &= \beta_i + \beta_j + \sum_{k \neq i,j} \beta_k x_k \\
\Rightarrow P(\psi \mid x_i = 1, x_j = 1) &\simeq \mathcal{N}(\psi; \beta_i + \beta_j, \sigma_B), \ \sigma_B = \sum_{i=1}^n \beta_i^2
\end{split}
\end{equation}
The expected reward for factor $i$ conditioned on $x_i=1, x_j=1$: 
\begin{equation}
\begin{split}
(z_i \mid x_i=1, x_j=1) &= \beta_i + \beta_j + \frac{1}{2}\sum_{k \neq i,j} \beta_k(1+x_k) \\
&= \frac{1}{2}(1 + \beta_i + \beta_j + \sum_{k \neq i,j}\beta_k x_k) \\
&= \frac{1}{2}(1 + \psi) \\
\Rightarrow \mathbb{E}(R_i \mid x_i=1, x_j=1) &= P(\psi > 0, \psi < 0) = 0
\end{split}
\end{equation}
The expected reward for factor $j$ conditioned on $x_i=1, x_j=1$: 
\begin{equation}
\begin{split}
(z_j \mid (x_i=1, x_j=1) &= \beta_i + \beta_j + \frac{1}{2}\sum_{k \neq i,j} \beta_k(1+x_k) \\
&= \frac{1}{2}(1 + \psi) \\
\Rightarrow \mathbb{E}(R_j \mid x_i=1, x_j =1) &= P(\psi > 0, \psi < 0) =0
\end{split}
\end{equation}
The expected reward for factor $i$ conditioned on $x_i=1, x_j=1$. In considering factor $l$ we must separately consider the two cases where $x_l = 1$ and $x_l = -1$: 
\begin{equation}
\begin{split}
(z_l \mid x_i=1, x_j=1, x_l = 1) &= \beta_i + \beta_j + \beta_l + \frac{1}{2}\sum_{k \neq i,j,l} \beta_k(1+x_k) \\
&= \frac{1}{2}(1 + \psi) \\
\Rightarrow \mathbb{E}(R_l \mid x_i=1, x_j =1, x_l = 1) &= P(\psi > 0, \psi < 0) = 0 \\
(z_l \mid x_i=1, x_j=1, x_l = -1) &= \beta_l + \frac{1}{2}\sum_{k \neq i,j,l} \beta_k(1-x_k) \\
&= \frac{1}{2}(1 - \psi + 2\beta_l) \\
\Rightarrow \mathbb{E}(R_l \mid x_i=1, x_j =1, x_l = -1) &= P(\psi < 0, \psi - 2\beta_l > 0) = 0 \\
\therefore \mathbb{E}(R_l \mid x_i=1, x_j =1) &= 0
\end{split}
\end{equation} 
\textbf{In summary}: 
\begin{equation}
\begin{split}
\mathbb{E}(R_i \mid x_i=1, x_j=1) &=0 \\
\mathbb{E}(R_j \mid x_i=1, x_j=1) &=0 \\ 
\mathbb{E}(R_l \mid x_i=1, x_j=1) &=0
\end{split}
\end{equation}
\\\\
\noindent \textbf{Scenario 2}, $x_i = -1, x_j = -1$, following the same procedure:
The value of $\psi$:
\begin{equation}
\begin{split}
(\psi \mid x_i = -1, x_j = -1) &= -\beta_i - \beta_j + \sum_{k \neq i,j} \beta_k x_k \\
\Rightarrow P(\psi \mid x_i = -1, x_j = -1) &\simeq \mathcal{N}(\psi; -\beta_i - \beta_j, \sigma_B), \ \sigma_B = \sum_{i=1}^n \beta_i^2
\end{split}
\end{equation}
The expected reward for factor $i$ conditioned on $x_i=-1, x_j=-1$: 
\begin{equation}
\begin{split}
(z_i | x_i=-1, x_j=-1) &= \beta_i + \beta_j + \frac{1}{2}\sum_{k \neq i,j} \beta_k(1-x_k) \\
&= \frac{1}{2}(1 + \beta_i + \beta_j - \sum_{k \neq i,j}\beta_k x_k) \\
&= \frac{1}{2}(1 - \psi) \\
\Rightarrow \mathbb{E}(R_i \mid x_i=-1, x_j=-1) &= P(\psi < 0, \psi > 0) = 0
\end{split}
\end{equation}
The expected reward for factor $j$ conditioned on $x_i=-1, x_j=-1$: 
\begin{equation}
\begin{split}
(z_j \mid x_i = -1, x_j = -1)  &= \beta_i + \beta_j + \frac{1}{2}\sum_{k \neq i,j} \beta_k(1-x_k) \\
&= \frac{1}{2}(1 - \psi) \\
\Rightarrow \mathbb{E}(R_j \mid x_i=-1, x_j =-1) &= P(\psi < 0, \psi > 0) =0
\end{split}
\end{equation}
The expected reward for factor $l$ conditioned on $x_i=-1, x_j=-1$: 
\begin{equation}
\begin{split}
(z_l \mid x_i=-1,x_j=-1, x_l = 1) &= \beta_l + \frac{1}{2}\sum_{k \neq i,j, l} \beta_k(1+x_k) \\
&= \frac{1}{2}(1 + \psi) \\
\Rightarrow \mathbb{E}(R_l \mid x_i=-1, x_j =-1, x_l = -1) &= P(\psi > 0, \psi < 0) = 0 \\
(z_l \mid x_i=-1, x_j=-1, x_l = -1) &= \beta_i + \beta_j + \beta_l + \frac{1}{2}\sum_{k \neq i,j,l} \beta_k(1-x_k) \\
&= \frac{1}{2}(1 - \psi + 2\beta_l) \\
\Rightarrow \mathbb{E}(R_l \mid x_i=-1, x_j =-1, x_l = -1) &= P(\psi < 0, \psi - 2\beta_l > 0) = 0 \\
\therefore \mathbb{E}(R_l \mid x_i=-1, x_j =-1) &= 0
\end{split}
\end{equation} 
\textbf{In summary}:
\begin{equation}
\begin{split}
\mathbb{E}(R_i \mid x_i=-1, x_j=-1) &=0 \\
\mathbb{E}(R_j \mid x_i=-1, x_j=-1) &=0 \\
\mathbb{E}(R_l \mid x_i=-1, x_j=-1) &=0
\end{split}
\end{equation}
\\\\
\noindent \textbf{Scenario 3}, $x_i = 1, x_j = -1$, following the same procedure:
The value of $\psi$:
\begin{equation}
\begin{split}
(\psi \mid x_i = 1, x_j = -1) &= \beta_i - \beta_j + \sum_{k \neq i,j} \beta_k x_k \\
\Rightarrow P(\psi \mid x_i = 1, x_j = -1) &\simeq \mathcal{N}(\psi; \beta_i - \beta_j, \sigma_B), \ \sigma_B = \sum_{i=1}^n \beta_i^2
\end{split}
\end{equation}
The expected reward for factor $i$ conditioned on $x_i=1, x_j=-1$: 
\begin{equation}
\begin{split}
(z_i | x_i=1, x_j=-1) &= \beta_i + \Delta + \frac{1}{2}\sum_{k \neq i,j} \beta_k(1+x_k) \\
&= \frac{1}{2}(1 + \beta_i -\beta_j + 2\Delta + \sum_{k \neq i,j}\beta_k x_k) \\
&= \frac{1}{2}(1 +\psi + 2\Delta) 
\Rightarrow \mathbb{E}(R_i \mid x_i=1, x_j=-1) = P(\psi > 0, \psi + 2\Delta < 0) = 0
\end{split}
\end{equation}
The expected reward for factor $j$ conditioned on $x_i=1, x_j=-1$: 
\begin{equation}
\begin{split}
(z_j \mid x_i = 1, x_j = -1)  &= \beta_j -\Delta + \frac{1}{2}\sum_{k \neq i,j} \beta_k(1-x_k) \\
&= \frac{1}{2}(1 - \psi -2\Delta) \\
\Rightarrow \mathbb{E}(R_j \mid x_i=1, x_j =-1) &= P(\psi < 0, \psi+2\Delta > 0) \simeq 2\mathcal{N}(\beta_i-\beta_j, 0, \sigma_B)\Delta
\end{split}
\end{equation}
The expected reward for factor $l$ conditioned on $x_i=1, x_j=-1$: 
\begin{equation}
\begin{split}
(z_l \mid x_i=1,x_j=-1, x_l = 1) &= \beta_i + \Delta+  \beta_l + \frac{1}{2}\sum_{k \neq i,j, l} \beta_k(1+x_k) \\
&= \frac{1}{2}(1 + \psi + 2\Delta) \\
\Rightarrow \mathbb{E}(R_l \mid x_i=1, x_j =-1, x_l = 1) &= P(\psi > 0, \psi +2\Delta < 0) = 0 \\
(z_l \mid x_i=-1, x_j=1, x_l = -1) &= \beta_j -\Delta + \beta_l + \frac{1}{2}\sum_{k \neq i,j, l} \beta_k(1-x_k) \\
&= \frac{1}{2}(1 - \psi - 2\beta_l) \\
\Rightarrow \mathbb{E}(R_l \mid x_i=1, x_j =-1, x_l = -1) &= P(\psi < 0, \psi + 2\beta_l > 0) \simeq 2\mathcal{N}(\beta_i-\beta_j+\beta_l, 0, \sigma_B)\Delta \\
\therefore \mathbb{E}(R_l \mid x_i=1, x_j =-1) &\simeq \mathcal{N}(\beta_i-\beta_j+\beta_l, 0, \sigma_B)\Delta
\end{split}
\end{equation} 
\textbf{In summary}: 
\begin{equation}
\begin{split}
\mathbb{E}(R_i \mid x_i=1, x_j=-1) &=0 \\
\mathbb{E}(R_j \mid x_i=1, x_j=-1) &\simeq 2 \Delta \mathcal{N}(\beta_i-\beta_j, 0, \sigma_B) \\
\mathbb{E}(R_l \mid x_i=1, x_j=-1) &\simeq \Delta \mathcal{N}(\beta_i-\beta_j+\beta_l, 0, \sigma_B)
\end{split}
\end{equation}
\\\\
\noindent \textbf{Scenario 4}, $x_i = -1, x_j = 1$, following the same procedure:
The value of $\psi$:
\begin{equation}
\begin{split}
(\psi \mid x_i = -1, x_j = 1) &= -\beta_i + \beta_j + \sum_{k \neq i,j} \beta_k x_k \\
\Rightarrow P(\psi \mid x_i = -1, x_j = 1) &\simeq \mathcal{N}(\psi; -\beta_i + \beta_j, \sigma_B), \ \sigma_B = \sum_{i=1}^n \beta_i^2
\end{split}
\end{equation}
The expected reward for factor $i$ conditioned on $x_i=-1, x_j=1$: 
\begin{equation}
\begin{split}
(z_i | x_i=-1, x_j=1) &= \beta_i + \Delta + \frac{1}{2}\sum_{k \neq i,j} \beta_k(1-x_k) \\
&= \frac{1}{2}(1 + \beta_i -\beta_j + 2\Delta - \sum_{k \neq i,j}\beta_k x_k) \\
&= \frac{1}{2}(1 - \psi + 2\Delta) 
\Rightarrow \mathbb{E}(R_i \mid x_i=-1, x_j=1) = P(\psi < 0, \psi - 2\Delta > 0) = 0
\end{split}
\end{equation}
The expected reward for factor $j$ conditioned on $x_i=-1, x_j=1$: 
\begin{equation}
\begin{split}
(z_j \mid x_i = -1, x_j = 1)  &= \beta_j -\Delta + \frac{1}{2}\sum_{k \neq i,j} \beta_k(1+x_k) \\
&= \frac{1}{2}(1 + \psi -2\Delta) \\
\Rightarrow \mathbb{E}(R_j \mid x_i=-1, x_j =1) &= P(\psi > 0, \psi-2\Delta < 0) \simeq 2\mathcal{N}(\beta_i-\beta_j, 0, \sigma_B)\Delta
\end{split}
\end{equation}
The expected reward for factor $l$ conditioned on $x_i=-1, x_j=1$: 
\begin{equation}
\begin{split}
(z_l \mid x_i=-1,x_j=1, x_l = 1) &= \beta_j -\Delta \beta_l + \frac{1}{2}\sum_{k \neq i,j, l} \beta_k(1+x_k) \\
&= \frac{1}{2}(1 + \psi - 2\Delta) \\
\Rightarrow \mathbb{E}(R_l \mid x_i=-1, x_j =1, x_l = 1) &= P(\psi > 0, \psi -2\Delta < 0) \simeq 2\mathcal{N}(\beta_i-\beta_j-\beta_l, 0, \sigma_B)\Delta\\
(z_l \mid x_i=-1, x_j=1, x_l = -1) &= \beta_i + \Delta + \beta_l + \frac{1}{2}\sum_{k \neq i,j, l} \beta_k(1-x_k) \\
&= \frac{1}{2}(1 - \psi + 2\beta_l) \\
\Rightarrow \mathbb{E}(R_l \mid x_i=-1, x_j =1, x_l = -1) &= P(\psi < 0, \psi - 2\beta_l > 0) = 0 \\
\therefore \mathbb{E}(R_l \mid x_i=-1, x_j =1) &\simeq \mathcal{N}(\beta_i-\beta_j-\beta_l, 0, \sigma_B)\Delta
\end{split}
\end{equation} 
\textbf{In summary}:
\begin{equation}
\begin{split}
\mathbb{E}(R_i \mid x_i=-1, x_j=1) &=0 \\
\mathbb{E}(R_j \mid x_i=-1, x_j=1) &\simeq 2  \mathcal{N}(\beta_i-\beta_j, 0, \sigma_B)\Delta  \\
\mathbb{E}(R_l \mid x_i=-1, x_j=1) &\simeq \mathcal{N}(\beta_i-\beta_j-\beta_l, 0, \sigma_B)\Delta  
\end{split}
\end{equation}
\\\\
\noindent Adding all four scenarios together weighted by the probability $p=1/4$ of each, we therefore have overall
\begin{equation}
\begin{split}
\mathbb{E}(R_i) &=0 \\
\mathbb{E}(R_j) &\simeq  \mathcal{N}(\beta_i-\beta_j, 0, \sigma_B)\Delta\\
\mathbb{E}(R_l) &\simeq \frac{1}{4}  \mathcal{N}(\beta_i-\beta_j+\beta_l, 0, \sigma_B)\Delta + \frac{1}{4}  \mathcal{N}(\beta_i-\beta_j-\beta_l, 0, \sigma_B)\Delta\\
&\simeq \frac{1}{2} \mathcal{N}(\beta_i-\beta_j, 0, \sigma_B)\Delta , \ \textrm{based on Taylor expansion for small } \beta_l
\end{split}
\end{equation} 
If we substitute these quantities into the replicator equation we obtain:
\begin{equation}
\begin{split}
\dot{\rho_i} &= \dot{\Delta} \simeq  -\frac{1}{2}\mathcal{N}(\beta_i-\beta_j, 0, \sigma_B)\Delta \\
\dot{\rho_j} &= -\dot{\Delta} \simeq \frac{1}{2}  \mathcal{N}(\beta_i-\beta_j, 0, \sigma_B)\Delta \\
\dot{\rho_l} &= 0
\end{split}
\end{equation}
Hence the initial perturbation will move back towards zero on both factors $i$ and $j$ without causing additional perturbations on the other factors, with the rate of convergence depending on the difference between $\beta_i$ and $\beta_j$
\\\\
\noindent \textbf{Perturbation on many factors}. Now we consider an extensive pertubation to the ideal distribution, such that $\rho_i = \beta_i + \Delta_i$. Specifically we assume that the values of $\{\Delta\}$ are sufficiently well spread across factors that the quantity $\sum_{k=1}^n \Delta_k x_k$ obeys the Lindeberg condition \cite{feller1970prob} and thus can be approximated as converging to a normal distribution for large $n$. Considering the case where the focal player observes $x_i=1$ without loss of generality, this implies that the expected reward for attending to factor $i$ can be written as:
\begin{equation}
\mathbb{E}(R_i) = P(\psi > 0, z_i < 1/2 | x_i = 1)
\end{equation}
Define $u \equiv 2z_i-1 = \psi + \Delta_i + \sum_{j \neq i} \Delta_j x_j)$, such that the condition above becomes:
\begin{equation}
\begin{split}
\mathbb{E}(R_i) &= P(\psi > 0, u_i < 0 | x_i = 1) \\
&= \int_{-\infty}^0 \int_0^\infty p(u_i \mid \psi) p(\psi \mid x_i =1)d\psi du
\end{split}
\end{equation}
In the limit of large $n$ we have the following limiting normal distributions:
\begin{equation}
\begin{split}
P(\psi | x_i=1) &= \mathcal{N}(\psi; \beta_i, \sigma_B), \ \sigma_B^2 = \sum_{j=1}^n \beta_j^2 \\
P(u_i | \psi) &= \mathcal{N}(u_i; \psi + \Delta_i, \sigma_\Delta), \ \sigma_\Delta^2 = \sum_{j=1}^n \Delta_j^2 \label{eqn:stability_norm_limits}
\end{split}
\end{equation}
In order to facilitate taking the limit as the perturbations $\{\Delta\}$ tend to zero, define $\Delta_i = k\delta_i$ and $\sigma_\delta^2 = \sum_{j=1}^n \delta_i^2 = k^2 \sigma_\Delta^2$, where $\delta_i$ characterises the relative sizes of the perturbations on each factor and $k$ an overall scale of perturbation. With these definitions we can write the expected reward as:
\begin{equation}
\mathbb{E}(R_i) = \int_0^\infty \mathcal{N}(\psi; \beta_i, \sigma_B) \int_{-\infty}^0 \mathcal{N}(u; \psi + k\delta_i, k\sigma_\delta) du d\psi \label{eqn:expected_reward_k_form}
\end{equation}
Making a change of variables of $w = \frac{-(\psi + k\delta_i)}{k\sigma_\delta}$, dropping terms of order $k^2$ and replacing one integral by the cumulative normal function in equation \ref{eqn:expected_reward_k_form} we get:
\begin{equation}
\begin{split}
\mathbb{E}(R_i) &\simeq k\sigma_\delta \mathcal{N}(\beta_i; 0, \sigma_B) \int_{-\infty}^{-\delta_i/\sigma_\delta} \Phi(w) dw \\
&= k\mathcal{N}(\beta_i; 0, \sigma_B)\left(-\delta_i\Phi(\frac{-\delta_i}{\sigma_\delta}) + \sigma_\delta \mathcal{N}(\frac{-\delta_i}{\sigma_\delta}; 0, 1)\right)
\end{split}
\end{equation}
We are now in a position to recognise that,since the values of $\{\Delta\}$ obey the Lindeberg conditon \cite{feller1970prob} by prrior assumption, as $n \rightarrow \infty$, $\frac{-\delta_i}{\sigma_\delta} \rightarrow 0$ and $\frac{\beta_i}{\sigma_B} \rightarrow 0$, which implies $\Phi(\frac{-\delta_i}{\sigma_\delta}) \simeq \frac{1}{2}$, $\mathcal{N}(\frac{-\delta_i}{\sigma_\delta}; 0, 1) \simeq \frac{1}{\sqrt{2\pi}}$ and $\mathcal{N}(\beta_i; 0, \sigma_B) \simeq \frac{1}{\sigma_B \sqrt{2\pi}}$. Substituting back in $\Delta_i = k\delta_i$ and $\sigma_\Delta = k \sigma_\delta$ the above then simplifies to
\begin{equation}
\mathbb{E}(R_i) \simeq \frac{1}{\sqrt{2\pi}}\left(-\frac{\Delta_i}{2} + \frac{\sigma_\Delta}{\sqrt{2\pi}} \right)
\end{equation}
The evolution of the density of individuals attending to each factor is determined by the relative expected rewards for that factor compared to the average expected reward across all factors, implying the following evolutionary equations for the perturbation after dropping second order terms:
\begin{equation}
\begin{split}
\dot{\rho_i} &= \rho_i \left( \mathbb{E}(R_i) - \sum_{j=1}^n \rho_j \mathbb{E}(R_j) \right) \\
\Rightarrow \dot{\Delta_i} &= \frac{\beta_i}{2\sqrt{2\pi}\sigma_B}\left(-\Delta_i + \sum_{j=1}^n \beta_j \Delta_j \right) \label{eqn:pertubation_evolution}
\end{split}
\end{equation}
This equation shows that the ideal distribution is generally stable to perturbations, though not absolutely so in the first order. If the perturbation values $\{\Delta\}$ are independent the values of $\{\beta\}$, the final term $\sum_{j=1}^n \beta_j \Delta_j$ tned to zero as $n$ tends to infinity. However, if the perturbation is dependent on $\{\beta\}$, this second term could cause the perturbation on some factors to grow for a short time. Once the perturbation has grown sufficiently, the first term will bring it back towards the stationary state.

Our analysis has been unable to fully treat pertubations on many factors where these do not satisfy the Lindeberg condition \cite{feller1970prob} to permit treatment by normal distribution approximations. However, the convergence of our simulations to a stationary equilibrium close the ideal distribution of $\rho_i = \beta_i$ from multiple starting conditions suggests that this stationary point is stable for practical purposes. 

\section*{Correlated factors}
So far we have considered the values of $x_1, \ldots, x_n$ to be uncorrelated and independent of one another. However, in real world examples this assumption will frequently be violated, either because there are observable information sources that display intrinsic correlations (consider the various news sources with either a left-leaning or right-leaning political bias), or because individuals may attend to \emph{composite} sources: an individual may attend to multiple different information sources and create their own composite factor through some combination of these. To illustrate, an individual who reads multiple news sources may vote based on their own weighted average of the opinions expressed in these. News aggregation services, or professional analysts may produce similar composite factors that individuals can choose to attend to, similarly to the way in which one may invest in a mutual fund with a manager who picks investments for the fund as a whole.

We can show that under the minority reward system, the ideal distribution of $\rho_i \propto \beta_i$ remains stationary and stable if we allow values of $x_1, \ldots, x_n$ to exhibit correlations. The intrisinsic stationarity of the ideal distribution remains, since under this distribution the collective vote is always correct, implying that the probability to receive any reward is zero independent of which factor an individual attends to. To assess the stability of this distribution we proceed from modified versions of the expressions in equation \ref{eqn:stability_norm_limits}, accounting the the correlations between values of $x_1, \ldots, x_n$. Defining $q_{ij} = <x_ix_j>$ to be the covariance between factors $i$ and $j$, we have the following limiting normal distributions:
\begin{equation}
\begin{split}
p(\psi | x_i=1) &= \mathcal{N}(\psi; \beta_i, \sigma_B'), \ \sigma_B'^2 = \sum_{j=1}^n\sum_{l=1}^n \beta_j \beta_l q_{jl}\\
p(u_ | \psi, x_i=1) &= \mathcal{N}(u_i; \psi + \Delta_i, \sigma_\Delta'), \ \sigma_\Delta'^2 = \sum_{j=1}^n \sum_{l=1}^n \Delta_j \Delta_l q_{jl}
\end{split}
\end{equation}
From this point we can repeat the analysis from equation \ref{eqn:expected_reward_k_form} to equation \ref{eqn:pertubation_evolution}, taking the limit as the perturbation size tends to zero and as $n$ tends to infinity as before. Taking these limits will be valid when the effective number of independent factors tends to infinity in line with $n$, such that $\frac{-\delta_i}{\sigma_\delta'} \rightarrow 0$ and $\frac{\beta_i}{\sigma_B'} \rightarrow 0$  as $n \rightarrow \infty$, but not otherwise. To illustrate by a simple example: if $q_{ij} = 1 \, \forall i,j$ then there is effectively only one independent factor, and the arguments made for the limiting behaviour as $n \rightarrow \infty$ will clearly not hold. The ideal distribution will display the same stability for cases with correlated factors as for those with uncorrelated factors as the determinant of the covariance matrix $\textbf{q}$ tends to infinity, i.e. when the aggregate correlation between factors grows more slowly than the number of factors.

\end{document}